\documentclass[conference]{IEEEtran}
\IEEEoverridecommandlockouts
\usepackage{cite}
\usepackage{amsmath,amssymb,amsfonts}
\usepackage[linesnumbered,ruled]{algorithm2e}
\usepackage{graphicx}
\usepackage{eurosym}
\usepackage[top=54pt,bottom=54pt,left=54pt,right=54pt,heightrounded]{geometry}
\usepackage{url}
\usepackage{textcomp}
\begin{document}
%\newgeometry{top=72pt, bottom=54pt, right=54pt, left=54pt}

\title{dockChain: A Solution for Electric Vehicles Charge Point Anxiety\\
\thanks{J. O'Connell, B. Cardiff and R. Shorten are with University College Dublin, School of Electrical, Electronic and Communications Engineering, Belfield, Dublin 4, Ireland. Emails: julia.o-connell@ucdconnect.ie, barry.cardiff@ucd.ie, robert.shorten@ucd.ie}
}
\author{Julia O'Connell, Barry Cardiff and Robert Shorten}
\maketitle

\begin{abstract}
This paper addresses Charge Point Anxiety surrounding electric vehicles (EVs), an issue preventing the mass adoption of this greener mode of transport. We discuss the design and implementation of a charge point adapter called \textit{dockChain} that will help mitigate Charge Point Anxiety. The key feature of the dockChain is that it allows multiple EVs to connect simultaneously to a single charge by connecting the adapters together in a chain resulting in additional charging opportunities - without the need for infrastructural changes. We describe the operation of the network of adapters, the hardware components and charging policies for the adapter. A distributed algorithm that can detect the length of the chain in a dockChain network is also presented.
\end{abstract}

\section{Introduction}
Two of the most disruptive contemporary trends evident in society are collaborative consumption ({\em the sharing economy}) and the electrification of mobility. People are beginning to realise that the excess consumption of goods, and intermittent utilisation of these goods, is directly linked to pollution and wastefulness~\cite{capitalism}. Collaborative consumption is the term used to describe a new business model where consumers move from a sole to shared ownership model in order to reduce waste. Sharing and sustainability are connected concepts; many people who decide to adopt sharing practices consider their choices as being `better for the environment'~\cite{collab}. This has led to the emergence of many peer-to-peer sharing platforms, such as Airbnb, Sharing Vehicles, Bike Sharing, and many others.\newline  

%The way in which we view our world has changed profoundly in recent years as we become more aware of how the past decades of hyper-consumerism are affecting the planet negatively. Hyper-consumerism is the consumption of goods for non-functional purposes. Goods are often seen as status symbols and are bought not for use but for wealth display. People are beginning to realise that this excess of useless goods is directly linked to pollution and wastefulness~\cite{capitalism}. The opposite mentality is collaborative consumption, where consumers swap and share goods in order to reduce consumption. Sharing and sustainability are connected concepts; many people who decide to adopt sharing practices consider their choices as being `better for the environment'~\cite{collab}. This has led to the emergence of peer-to-peer sharing platforms, also known as the Sharing Economy. Examples of this business model are Airbnb, Sharing Vehicles, Bike Sharing, and many others. 

Mobility is one application domain where the sharing economy is having a real impact. In particular, shared mobility platforms such as car2go and DriveNow are changing the manner in which we own and use cars. However, despite these successes, structural impediments are hindering the evolution of collaborative consumption based mobility business models in our cities. This is particularly true in the domain of vehicle electrification where vehicle electrification and consumer behaviour often do not facilitate sharing economy ideas. It is in this context that this paper is presented. Specifically, our objective is to present an infrastructure component to facilitate the adoption of collaborative consumption ideas in the context of electric vehicles.\\

Our motivation for doing this is evident.;  in 2015, every country in the world had signed the Paris Agreement\footnote{The U.S. declared in mid 2017 that it will pull out of this agreement but this has not yet happened} that pledges to reduce greenhouse emission gases (GHG) by 40\% by 2030~\cite{paris}. The transport sector alone is responsible for almost one quarter of European greenhouse emission gases (GHG) and 72.9\% of this is due to road transportation~\cite{greenhouse}. This pressure has brought about the support of greener sources of transport that reduce the pollution of GHG, namely Electric Vehicles. As a result, it is foreseen that electric vehicles will become the major form of road transportation in the coming years.  Many countries have also set aggressive targets for phasing out petrol and diesel cars and increasing the number of EVs on the road\footnote{`A Brighter Future for Electric Cars and the Planet' by The Editorial Board, The New York Times, 2017}. In the UK, the proposed ban on sale of petrol and diesel cars has been brought forward by a decade to 2030\footnote{`National Grid backs plan for earlier petrol and diesel ban' by Adam Vaughan, The Guardian, March 2018}. The number of PEVs in use worldwide has almost doubled every year since 2012, from 113,000 in 2012 to approximately 1.209 million in 2016~\cite{no-ev}. A similar trend can be seen in Ireland, in Table~\ref{table:numev}, with the number of EVs on the road increasing steadily, particularly in 2016\footnote{2017 data has not yet been released}.

\setlength{\intextsep}{0pt}
\setlength{\textfloatsep}{0pt}
\begin{table}[!htbp]
\centering
\caption{Number of EVs and Hybrid Electric Vehicles (HEVs) Licensed for the First Time in Ireland~\cite{cso}}
\label{table:numev}
\begin{tabular}{ |c | c || c | c|}
	\hline
	Year & No. EV Licensed & Year & No. EV Licensed \\ [1ex]
	\hline
	\hline
	2009 & 351 & 2013 & 654 \\
	2010 & 805 & 2014 & 1,233\\
	2011 & 661 & 2015 & 1,899 \\
	2012 & 850 & 2016 & 2,982\\
	\hline
\end{tabular}
\end{table}

Yet, despite this increase, the adoption of EVs as a proportion of the overall transport sector has been disappointing~\cite{demand}, with only 1.1\% of all vehicles sold globally in 2016 being EVs~\cite{energy-agency}. A good example of this shortfall in growth compared to expectations can be seen in Ireland where there are currently approximately 9,000 EVs and HEVs (Hybrid Electric Vehicles) in use~\cite{cso}, but the government had set ambitious target to have 230,000 EVs on the road by 2020 - a figure that was subsequently revised in 2017 downwards to 20,000\footnote{`Ireland leading a weak charge on electric vehicles' by Kevin O?Sullivan, The Irish Times, July 2017}. This hesitation to further adopt electric vehicles can be narrowed down to a few \emph{perceived} issues that consumers may have. The main documented problems regarding EVs are: battery related issues, cost, range anxiety, vehicle size, long charging times and electromagnetic emissions~\cite{ev-opt}. Many of these issues are being resolved, for example:\\
\begin{enumerate}
	\item[1] \emph{Cost:} By comparison to equivalent Internal Combustion Engine (ICE) vehicles EVs tend to have higher upfront costs due in part to the high cost of lithium ion batteries~\cite{battery}. Many governments offer subsidies, for example, in Ireland the Sustainable Energy Authority of Ireland (SEAI) offers subsidies of up to \euro5,000 towards the purchase of an EV~\cite{grant}. A forecast by Bloomberg New Energy Finance has stated that EVs will become cheaper than conventional cars \textit{without} government subsidies by 2025 to 2030\footnote{`The Electric Car Revolution is Accelerating' by Jess Shankleman, Bloomberg, July 2017}. That being the case cost will no longer be a barrier to the adoption of EVs.\\
	\item[2] \emph{Range Anxiety:} Another well documented issue is that of {\em range anxiety}; namely,  the fear that occurs that there may not be sufficient range to cover the drivers' need or that EV owners. The range of EVs is increasing dramatically every year. The maximum all-electric vehicle range has grown from 151km (94 miles) in 2011 to 540km (335 miles) in 2017~\cite{range-us}, which is greater than the average range of an ICE, of 400km - 500km per full tank of petrol/diesel. Thus, range anxiety is also no longer a real barrier for the adoption of EVs.\\
\end{enumerate}
Despite this progress some challenges remain. Insufficient public infrastructure is still cited as the biggest impediment to the adoption of EVs, however, contrary to popular belief, this s not an issue many countries. For example, in Ireland, where there are approximately 2500 plug-in EVs in use with 1200 public charge points\footnote{ESB ecars}. The real problem is a consumer behaviour related issue leading a phenomenon called Charge Point Anxiety - it is this problem that is addressed in this paper.
\begin{enumerate}
	\item[3]  \emph{Charge Point Anxiety:}  During business hours charging spaces tend to be fully occupied in the many urban centres either by EV owners parking there for the entire workday despite the EV being fully charged within 2 hours, or by ICE vehicles illegally occupying designated spaces. In either scenario the charge point is unavailable to other users for large portions of the day resulting in under-utilization of valuable infrastructure and EV car owners experiencing Charge Point Anxiety; this is one of the main barriers preventing the mass uptake of EVs.\\
\end{enumerate}

The objective of this paper is to address Charge Point Anxiety through the development of the `dockChain' adapter. The most intuitive solution to charge point anxiety would be increase the number of charge points in these areas but, unfortunately, this is not feasible. The cost of building a single AC public charge point can range from \$600 - \$12,660~\cite{cost}. It is expected that the phenomenon of Charge Point Anxiety will become worse as adoption of EVs outstrips infrastructure.  Our basic idea is to develop an adapter is to extend the reach of charge points to allow multiple EVs to connect simultaneously. To this end we designed the dockChain adapter that allows the adapters to connect in a `daisychained' or `cascaded' manner as shown below in Figure~\ref{fig:chain}. Specifically, each EV owner connects to the charge point with their adapter and connects their car to that adapter.\newline 

{\bf Comment :} It is important to note that the inability to connect vehicles to the network is not just an inconvenience for EV owners. It also reduces the ability of the electrical grid to store energy in the EV fleet. This is an important consideration in the design of grid ancillary services - especially for V2G services and in using the vehicle fleet as a storage buffer.\newline 

{\bf Comment :} Before proceeding it is important to note that the present adaptor builds on our previous work, in particular, a `SmartPlug' that aimed to make private charge points public, enable dual charging and allow home owners to monetise their unused asset~\cite{cp-anxiety}. However, it is important to note that the focus of the current adaptor is to extend the availability of public charge points - rather than to monetise private ones.\newline

\setlength{\intextsep}{0pt}
\setlength{\textfloatsep}{0pt}
\begin{figure}[htbp]
\centerline{\includegraphics[scale = 0.18]{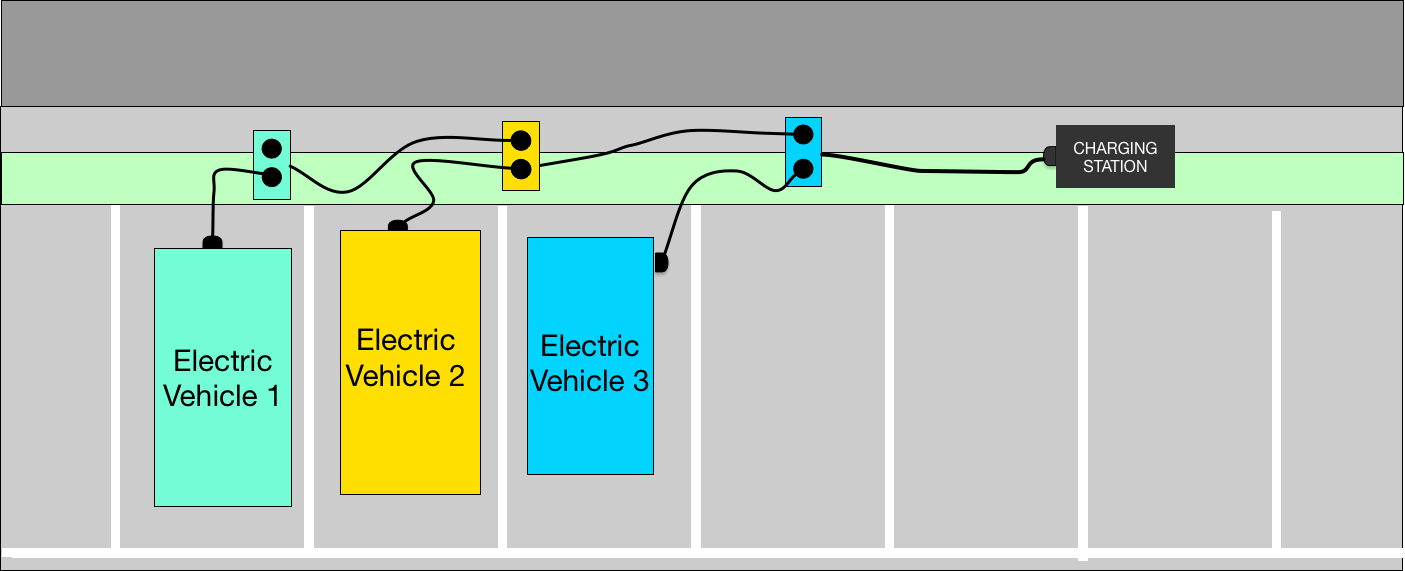}}
\caption{Three vehicles connected to a single charge point and charging simultaneously using three dockChain adapters}
\label{fig:chain}
\end{figure}
This paper is structured as follows. In Section~\ref{sec:the_dockChain_adaptor} we describe the design of the operation of the dockChain adaptor, how to set it up and how it can handle vehicles and adapters being added and/or removed from the chain. In Section~\ref{main:hardware}, the main hardware components are described. In Section~\ref{main:policies}, we present the various charging policies that can be implemented giving weighting of priority to different vehicles based on different criteria. Finally in Section~\ref{main:algo} \&~\ref{main:results}, an algorithm to detect the number of EVs connected at any one time is introduced, simulated and analysed.

\setlength{\intextsep}{0pt}
\setlength{\textfloatsep}{0pt}
\begin{figure*}
	\centering\includegraphics[width=.6\linewidth,height=2.7in]{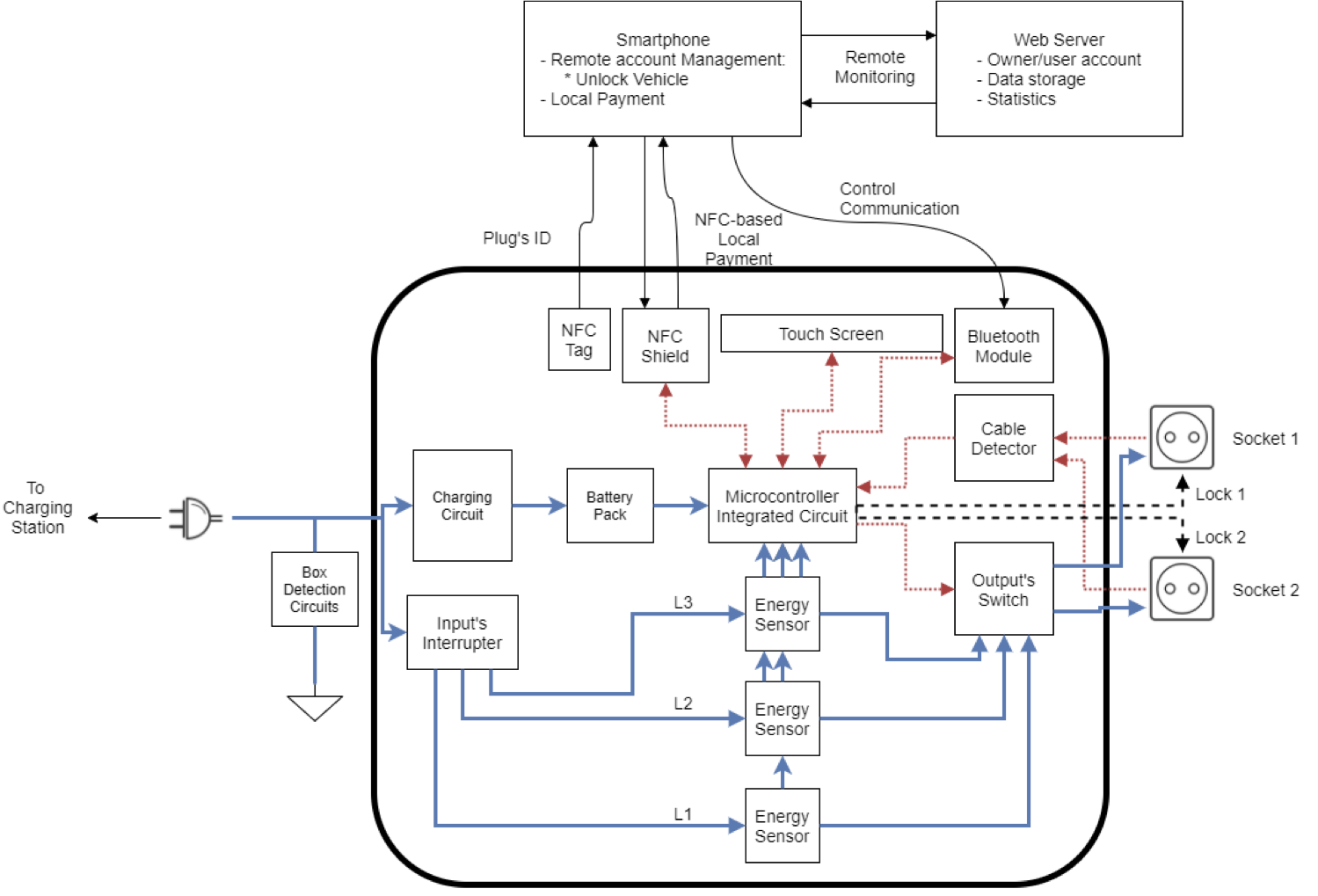}\par
	\caption{Schematic of dockChain adapter}
	\label{fig:schematic}
\end{figure*}

\section{The dockChain adaptor}\label{sec:the_dockChain_adaptor}
The dockChain adapter enables manyfold access to a single charge point. It also allows the adapters to be connected together in a `cascading' fashion to create a network that enables multiple vehicles to be charged simultaneously. The `cascading' feature is a key innovation for this adapter. It alleviates the feeling of Charge Point Anxiety as it can extend charge point accessibility. In particular, it prevents a single vehicle blocking the charge point. Also if the first EV owner who arrives at a charge point uses this adapter it allows another EV owner to arrive later and still be able to charge their vehicle.

It is envisaged that each car will carry a dockChain adaptor in much the same way that a spare wheel is carried. When connecting to a charge point, the car owner connects the adaptor to the charge point (or to another dockChain adaptor), then connects the car to his/her dockChain adaptor - as depicted in Figure 1. 

A single dockChain adapter allows dual charging of two EVs and can be used at a public charge point which allows safe three phase AC charging up to 32A. This adapter can also be used at private and semi-public charge points which are typically single-phase and three-phase respectively up to 16A.
 
The operation of the docChain adaptor is different depending on the scenario as follows:

\subsection{Adding dockChains and EV}
\subsubsection{First to Arrive with dockChain}
When an EV owner arrives at an empty charge point they plug the dockChain adapter into the charge point (it powers up automatically) and their EV into the adapter - these case be done in any order. The adaptor recognises the presence of both the adaptor and the EV and, through the use of a touch-screen interface, asks the user to confirm their desire to proceed. Upon confirmation the adaptor begins the charging process and the EV is locked into place preventing unauthorized removal.

\subsubsection{Subsequent arrivals with dockChain}
An EV owner arriving at a charge point that already has a dockChain adapter with an empty socket may connect another dockChain adaptor to this empty socket following the same procedure as before. The end result will be two EVs connected to a sequence of dockChain boxes with one remaining free socket at the end. This process can continue again and again resulting in a chain $N$ adaptors long servicing $N$ EVs as illustrated in Figure \ref{fig:chain} for $N\text{=}3$.

\subsubsection{Subsequent arrivals without dockChain}
An EV owner arrives at a charge point that already has dockChain adapter (or chain of adaptors) connected and wishes to connect their EV into the empty socket. In this scenario the addition of this EV would, in effect, terminate the chain rendering it useless for any other (dockChain) users that may arrive at a later time. Thus the software on the dockChain, upon detection of an EV in the previously unused socket, notifies the user through the display that this EV will not received any change, and request that they un-plug their EV leaving it free for other dockChain users.
Importantly this EV will not be locked into place so other dockChain users arriving later can remove the offending EV and continue the dockChain as intended.
Another important point to note is that this policy of not allowing two EVs to connect to the same dockChain adaptors enforces a chain structure (not a binary tree), and ensures that there will always be exactly 1 empty socket at the end of the chain which will be used later in the chain length discovery protocol in Section \ref{main:algo}.

\subsection{Removal of dockChain and EV}
For theft prevention  reasons the removal of a dockChain and EV from a chain is protected by the use of a smartphone app that is one-time paired with an associated dockChain adaptor. It is through the use of this paired app that requests to unlock EVs are made to the adaptor over a bluetooth link.

\subsubsection{Removal from end of chain}
An EV owner returns to a chain and wishes to remove their dockChain adaptor and EV from the end of the chain. The user employs their smartphone app to request that dockchain release their EV allowing the EV to be removed. Additionally the adaptor then electrically disconnects itself on the upstream side (be it another dockChain adaptor or a actual charge point). This will cause the upstream device to also release the dockChain allowing it to be removed also thus completing the task.

\subsubsection{Remove from mid-chain} 
This the same as above except that at the end it is necessary (if possible) to reconnect the downstream dockChain adaptor. The adaptor being removed (and therefore the app) are aware of the presence of the downstream adaptor and so the user can be reminded to perform the reconnection on both the screen on the dockChain and on the smartphone UI. Moreover, each app can communicate its identity, status, actions and whereabouts to a online service that can in effect rank users' behaviour - a score that can be used to encourage good behaviour however this is beyond the scope of this paper. 
One comment is to note that, with all the best intentions, it will inevitably happen that sometime the cables are going to be just too short to make the connection and the chain will remain broken.

\section{Main Hardware Components}\label{main:hardware}

We now give a brief description of the hardware components of the dockChain adapter, as shown in Figure~\ref{fig:schematic}.

\begin{enumerate}
	\item \textit{Touchscreen Interface:} The interface for the adapter is a small touchscreen. This allows the user to start/stop charging for each socket. The interface also displays the state of both sockets, the availability and summary of the charging session.\newline
	\item \textit{Bluetooth:} Bluetooth is an integral part of the design. Bluetooth is used as the communication method between the user's smartphone and the adapter. The user can use an app to send a request to the adapter to unlock a vehicle.\newline
 	\item \textit{Microcontroller:} The operation of the adapter is coordinated by a Raspberry Pi 3 Model B microcontroller. It manages: (i) the monitoring of the plug-interface, i.e. listening for any user requests and showing the current state of the plug; (ii) the orchestration of the output signals depending on the state of the sockets; (iii) the monitoring of the energy consumed by each vehicle; and (iv) the communication between the adapter and the smartphone app.\newline
	\item \textit{Energy Sensor:} The energy being consumed by the dockChain adapter is measured on each of the three phases using a PZEM004T power meter.\newline 
	\item \textit{Electronic Cable Locks:} An electronic locking solenoid is used for each socket to secure the charging cables once the charging process has begun, and only will be released after a request from the user. This offers an advantage to the users of the adapter as it prevents the theft of the charging cable.\newline
\end{enumerate}

\section{Charging Policies}\label{main:policies}
The dockChain adaptor allows for a number of charging policies to be implemented. Some of these policies are introduced in the paper `On Charge Point Anxiety and the Sharing Economy'~\cite{cp-anxiety}. Examples include the following.\newline 
\begin{itemize}
	\item\textit{First come, First served: (water filling algorithm)} The first vehicle in the chain is charged first, then the second, and so forth.\newline
	\item\textit{Equal Charge:} Charging current is split equally among a vehicles in the chain. \newline
	\item\textit{Premium User:} If a user is willing to pay a `premium' price to give their vehicle a higher priority, then charging current is diverted accordingly.\newline
	\item\textit{Battery Level/Journey Length:} Priority given based on a vehicles battery level or the journey length.\newline
	\item\textit{Government Vehicles/ Carpool/ Business Vehicles:} Priority given based on if the vehicle is a government vehicle, a business vehicle or on the number of occupants of the vehicle.\newline
	\item\textit{Slot Trading:} Market based prioritisation based on trading of slots between vehicles for money.
\end{itemize}
For some of the above algorithms, it is important to know the number of cars connected to charge point via the dockChain network. We now discuss an algorithm that allows the this number to be determined in a disitributed manner. 
\section{Cascading and Distribution Algorithm}\label{main:algo}
The number of vehicles connected to the network must be determined in order to implement many of the charging policies mentioned in the previous section. Although all the adapters contain bluetooth capability and could potentially communicate the chain length through that medium, the motivation for developing this algorithm is twofold. First, there could be technical difficulties with bluetooth as the connectivity can be unreliable. Second, for trust and verification and to prevent dishonest EV owners from gaining an advantage.\newline
\begin{figure}[htbp]
\centerline{\includegraphics[scale = 0.3]{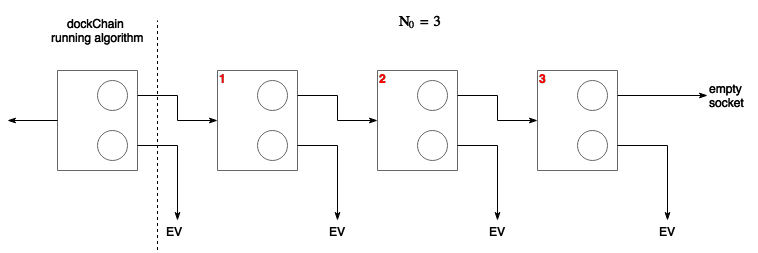}}
\caption{dockChain adapter running algorithm with a chain length of 3 adapters downstream, $N_0$ = 3.}
\label{fig:chain_length}
\end{figure}

{\bf Assumption 1 :} Before proceeding we note the following assumption. We assume that all EV owners are using the dockChain system (as depicted in Figure 2). More specifically, we assume that the last box in the chain always has one empty socket.\newline

{\bf Assumption 2 :} We also assume that each box in the chain runs the following algorithm, for a certain period of time, each time a car is disconnected, and the dockChain network is reconnected. Note that the adaptors can always detect this situation as: (a) they will momentarily lost power if they are left of the disconnecting car (ref. Figure 1); of (b) the charge distribution on their sockets will momentarily change if they are to the right of the disconnecting car.\newline   

{\bf Algorithm :} Each adapter randomly assigns current to a socket every $\tau$ secs for the algorithm duration of T = N$\tau$ with equal probability. Therefore, initially, the adapter offers power 50\% of the time to each socket. If an adapter has an empty socket then the current will not drain from the charge point. The current on each socket is measured and the fraction of the currend drawn is used to establish the number of cars attached to the socket.  The algorithm is depicted in Algorithm~\ref{alg:algo}.
\setlength{\intextsep}{0pt}
\setlength{\textfloatsep}{0pt}
\begin{algorithm}[!htbp]

    \SetKwInOut{Input}{Input}
    \SetKwInOut{Output}{Output}
    
    \Input{$p$, probability of choosing socket 0}
    \Output{Length of chain for socket 0 and socket 1 }
    
    Initially let  $p$ = 0.5 otherwise use input $p$\;
    Let socketUtilization = [0,0]\;
    
    \For{N}{
    	draw R.V, x,  $\in$ [0,1] with probability = $p$\\
	numberTimesSocket0Actitvated++\;
	provide charge to socket \#x\\
	measure current\\
	\If{current $>$ Threshold}{
		socketUtilization[x]++\;
	}
    }
    
    fractionUtilization[0] = socketUtilization[0] $\div$ numberTimesSocket0Activated\;
    fractionUtilization[1] = socketUtilization[1] $\div$ (N - numberTimesSocket0Activated)\;
  
    lengthOfChain[0] = $-\log_2$(1-fractionUtilization[0])\;
    lengthOfChain[1] = $-\log_2$(1-fractionUtilization[1])\;
\caption{Algorithm to detect number of vehicles attached to the network}
\label{alg:algo}
\end{algorithm}

\subsection{Chain Length Determination}
Each dockChain knows which socket is connected to an EV and which is connected to another dockChain. Let $N_o$ be the length of the chain downstream of a particular dockChain adaptor. Assuming that each dockChain operates an equal 50/50 allocation policy, then the empty socket at the end of the chain will we selected with probability $\frac{1}{2^{N_o}}$, and thus the utilisation on the $n^{th}$ activation is:
\begin{align}
	U_n =
	\begin{cases}
	0 ~~~\text{with probability }          \frac{1}{2^{N_o}}\\
	1 ~~~\text{with probability } 1\text{-}\frac{1}{2^{N_o}}
	\end{cases}
\end{align}
Thus $U_n$ is a Bernoulli distributed random variable implying that its sum over $N$ intervals is a Binomial distribution, and thus its average $\bar{U}_n$ has mean, $\mu$, and variance $\sigma^2$ given by:
\begin{align}
	\mu=p~~~~\text{and}~~~~\sigma^2=\frac{1}{N}p\left(1\text{-}p \right)
\end{align}
where $p \triangleq 1\text{-}\frac{1}{2^{N_o}}$

\begin{figure}[htbp]
\centerline{\includegraphics[scale = 0.64]{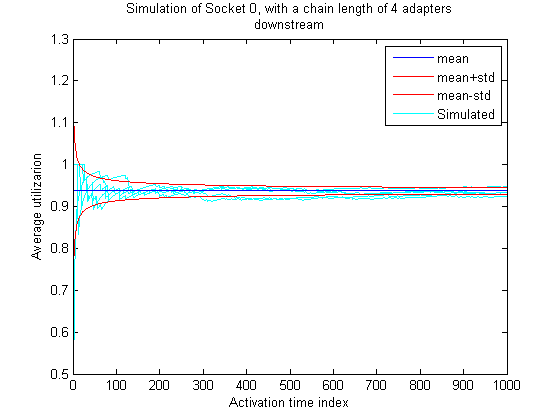}}
\setlength{\abovecaptionskip}{-5pt}
\setlength{\belowcaptionskip}{-5pt}
\caption{The average utilisation of Socket 0 over time plotted against the mean and standard deviation for a chain of 4 adapters with an empty socket. It converges to the expected value of 0.9375.}
\label{fig:avg_util}
\end{figure}

For illustration, the average utilisation of socket 0 with 4 adapters connected to it with one empty socket is simulated in Figure~\ref{fig:avg_util} for verification.

\setlength{\intextsep}{0pt}
\setlength{\textfloatsep}{0pt}

\subsection{`Equal Charge' Implementation}
Once the length of the chain is known, the amount of current allocated to each socket of a dockChain adaptor can be adapted to implement various policies.  For illustration, we now implement a policy to supply equal current to each car in the chain.  The following algorithm adjusts the probability value $p$ that is used as an input to Algorithm~\ref{alg:algo} based on the ratio of the length of the chain on each socket. The algorithm uses an IIR (Infinite Impulse Response) Filter to average the past values of p with the current ratio of chain lengths.  The algorithm is shown illustrated in Algorithm~\ref{alg:algo2}.

\setlength{\intextsep}{0pt}
\setlength{\textfloatsep}{0pt}
\begin{algorithm}
\linespread{1.35}\selectfont

    \SetKwInOut{Input}{Input}
    \SetKwInOut{Output}{Output}
    
     \Input{Length of Chain for socket 0 and socket 1}
     \Output{$p$, probability of choosing socket 0}
     
     \BlankLine
     p = ($\alpha$)p + $\frac{(1-\alpha)(lengthOfChain[0])}{lengthOfChain[0] + lengthOfChain[1]}$

\caption{To adapt probability p over time depending on the length of the chain in each socket}
\label{alg:algo2}
\end{algorithm}

\section{Algorithm Simulation Results}\label{main:results}
\begin{figure}
\centerline{\includegraphics[scale = 0.64]{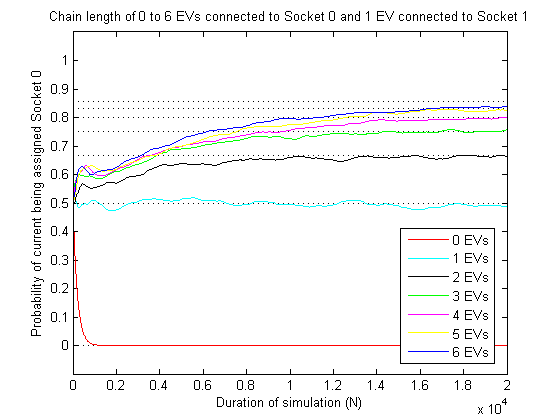}}
\setlength{\abovecaptionskip}{-5pt}
\setlength{\belowcaptionskip}{0pt}
\caption{The probability that current is assigned to Socket 0. Socket 0 chain length ranges from 0 EVs to 6 EVs and the Socket 1 chain length is 1 EV.}
\label{fig:1EV}
\end{figure}

\begin{figure}
\centerline{\includegraphics[scale = 0.64]{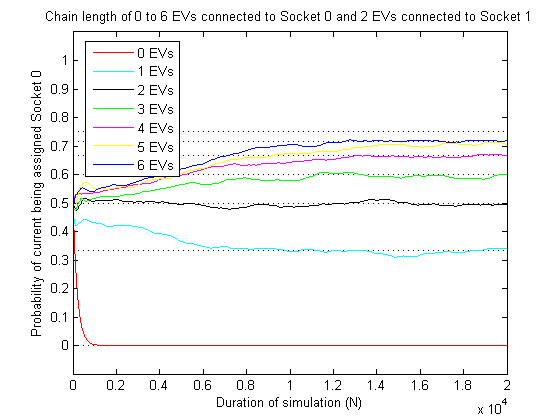}}
\setlength{\abovecaptionskip}{-5pt}
\setlength{\belowcaptionskip}{0pt}
\caption{The probability that current is assigned to Socket 0. Socket 0 chain length ranges from 0 EVs to 6 EVs and the Socket 1 chain length is 2 EVs.}
\label{fig:2EV}
\end{figure}

\begin{figure}
\centerline{\includegraphics[scale = 0.64]{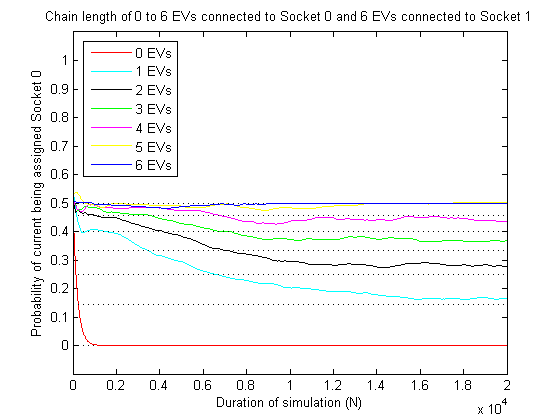}}
\setlength{\abovecaptionskip}{-5pt}
\setlength{\belowcaptionskip}{0pt}
\caption{The probability that current is assigned to Socket 0. Socket 0 chain length ranges from 0 EVs to 6 EVs and the Socket 1 chain length is 6 EVs.}
\label{fig:6EV}
\end{figure}

We now present simulation results to demonstrate the efficacy of Algorithm 1.  We chose six as the maximum length of chain as it is unlikely to encounter a chain longer than that in a real world situation. Socket 0 and 1 refer to the each of the sockets on the dockChain adapter. The simulated graphs shows the probability that current is assigned to Socket 0 based on the ratio of chain lengths, i.e. if the there is a chain length of 3 EVs on Socket 0 and 2 EVs connected to Socket 1 so the probability that current is assigned to Socket 0 is $\frac{3}{3+2} = 0.6$. The simulations are shown in Figure~\ref{fig:1EV} -~\ref{fig:6EV} with the dotted line showing the expected probability. The results from these simulations can be seen in Table~\ref{table:EV}. 

\setlength{\intextsep}{0pt}
\setlength{\textfloatsep}{0pt}
\begin{table}
\centering
\caption{Results of simulation regarding Figure~\ref{fig:1EV}-~\ref{fig:6EV}. Chain length of range of 0 to 6 EVs connected to Socket 0 and chain length of 1 EV, 2 EVs and 6 EVs connected to Socket 1.}
\begin{tabular}{ |c | c | c | c |}
	\hline
	Chain Length & Chain Length & Probability assigned & Simulated\\ [1ex]
	of Socket 0 & of Socket 1 & to Socket 0 & Probability\\
	\hline
	0 & 1 & 0 & 0\\
	1 & 1 & 0.5 & 0.52\\
	2 & 1 & 0.66 & 0.67\\
	3 & 1 & 0.75 & 0.745\\
	4 & 1 & 0.8 & 0.795\\
	5 & 1 & 0.83 & 0.84\\
	6 & 1 & 0.85 & 0.84\\
	\hline
	0 & 2 & 0 & 0\\
	1 & 2 & 0.33 & 0.326\\
	2 & 2 & 0.5 & 0.503\\
	3 & 2 & 0.6 & 0.6\\
	4 & 2 & 0.66 & 0.68\\
	5 & 2 & 0.714 & 0.712\\
	6 & 2 & 0.75 & 0.72\\
	\hline
	0 & 6 & 0 & 0\\
	1 & 6 & 0.143 & 0.164\\
	2 & 6 & 0.25 & 0.29\\
	3 & 6 & 0.33 & 0.36\\
	4 & 6 & 0.4 & 0.44\\
	5 & 6 & 0.46 & 0.49\\
	6 & 6 & 0.5 & 0.5\\
	\hline
\end{tabular}
\label{table:EV}
\end{table} 

\setlength{\intextsep}{0pt}
\setlength{\textfloatsep}{0pt}

\setlength{\intextsep}{0pt}
\setlength{\textfloatsep}{0pt}

\setlength{\intextsep}{0pt}
\setlength{\textfloatsep}{0pt}

\section{Conclusion}

In this paper, a charge point adapter, dockChain, was presented as a solution to Charge Point Anxiety, one of the main barriers preventing the mass adoption of EV that has yet to be solved. The operation of dockChain, its hardware components and possible charging policies that could be implemented have also been explored in more detail. We have presented two algorithms to help manage the `cascading' network feature of the adapter, i.e finding the length of the chain connected to each socket and redistributing the current equally to each vehicle connected to the network. 
\bibliography{conf}
\bibliographystyle{unsrt}

\end{document}